\documentstyle[11pt]{article}
\def\up#1{\raise 1ex\hbox{#1}}
\begin{document}
\centerline{\bf A REVIEW OF GROWING INTERFACES IN QUENCHED}
\centerline{\bf DISORDERED MEDIA}
\bigskip
L. A. Braunstein$^{\dagger}$\footnote{E-mail: lbrauns@mdp.edu.ar},
R. C. Buceta${^\dagger}$, A.
D\'{\i}az-S\'anchez$^{\ddagger}$\footnote{E-mail: tasio@fcu.um.es}
and N. Giovambattista$^{\S}$\par
\bigskip
$\dagger$ Departamento de F\'{\i}sica, Facultad de Ciencias
Exactas y Naturales,\par Universidad Nacional de Mar del Plata,
Funes 3350, (7600) Mar del Plata, Argentina.\par $\ddagger$
Departamento de F\'{\i}sica, Universidad de Murcia, E-30071
Murcia, Spain.\par $\S$ Center for Polymers Studies and
Department of Physics, Boston University,\par Boston,
Massachusetts 02215, USA.\par

\section*{Abstract}

We make a review of the two principal models that allows to
explain the imbibition of fluid in porous media. These models,
that belong to the directed percolation depinning (DPD)
universality class, where introduced simultaneously by the Tang
and Leschhorn [Phys. Rev A {\bf 45}, R8309 (1992)] and Buldyrev
{\sl et al.} [Phys. Rev. A {\bf 45}, R8313 (1992)] and reviewed by
Braunstein {\sl et al.} [J. Phys. A {\bf 32}, 1801 (1999); Phys.
Rev. E {\bf 59}, 4243 (1999)]. Even these models have been
classified in the same universality class than the
Kardar-Parisi-Zhang equation [Phys. Rev. Lett. {\bf 56}, 889,
(1986)] with quenched noise (QKPZ), the contributions to the
growing mechanisms are quite different. The lateral contribution
in the DPD models, leads to an increasing of the roughness near
the criticality while in the QKPZ equation this contribution
always flattens the roughness. These results suggest that the QKPZ
equation does not describe properly the DPD models even when the
exponents derived from this equation are similar to the one
obtained from the simulations of these models. This fact is
confirmed trough the deduced analytical equation for the Tang and
Leschhorn model. This equation has the same symmetries than the
QKPZ one but its coefficients depend on the balance between the
driving force and the quenched noise.

\section{Introduction}
\label{sec:intro}

In the last years there has been a growing interest in the
understanding a vast variety of scale invariant phenomena
occurring in nature. Experiments and observations indeed suggest
that many physical systems develop correlations with power law
behaviour both in space and time. However, the fact that certain
structures exhibit fractal and complex properties does not tell us
why this happens. Pattern formation, aggregation phenomena,
biological systems, geological systems, disordered materials,
clustering of matter in the universe and many fields in which
scale invariance has been observed as a common and basic feature.

A crucial point to understand is therefore the origin of the
general scale-invariance of natural phenomena. This would
correspond to the understanding of the origin of fractal
structures from the knowledge of the microscopic physical
processes at the basis of these phenomena.

In scale invariant phenomena, events and information spread over a
wide range of length and time scales, so that no matter what is
the size of the scale considered one always observes surprisingly
rich structures. These systems, with very many degrees of freedom,
are usually so complex that their large scale behaviour cannot be
predicted from the microscopic dynamics. New types of collective
behaviour arise and their understanding represents one of the most
challenging areas in modern statistical physics.

The concepts of scaling and power law behavior was introduced to
the study of critical phenomena in second order transitions. The
physics of complex systems, however, is new with respect to
critical phenomena. The theory of equilibrium statistical physics
is strongly based on the ergodic hypothesis and scale invariance
develops at the critical equilibrium between order and disorder.
Reaching this equilibrium requires the fine tuning of various
parameters. In usual critical phenomena the same exponents that
define the onset of magnetization also describe the liquid vapor
transition in water. This strong universality appears to be a
characteristic of equilibrium systems. On the contrary the origin
of the scale invariance in nature in the rich domain of
nonequilibrium systems is not so well understood. Systems far from
equilibrium, do not seem to exhibit the same degree of
universality as the fractal dimension can be easily altered by
simple changes in the growth process. This lack of universality is
sometimes viewed as a negative element because one is forced to
describe specific systems instead of a single universal model.
This is why such a problem can only be investigated using many
tools as computer simulations, analytical tools and suitably
designed experiments. While the theoretical activity is focused
mainly on Monte Carlo simulations, it is very important to
understand the relations between theory and real experiments.

Fractal geometry provided the mathematical framework for the
extension of these concepts to a vast variety of natural
phenomena. A principal subject where fractals play an essential
role is in the study of self-affine features arising in complex
systems with many degrees of freedom, such as of interface growth
in disordered systems including percolation properties of fluid
displacement in disordered media.

All these models have been extensively studied by computer
simulations. The information generated by these computer
experiments brings a visual intuition as a valuable tool in
science. In this respect computer simulations represent an
essential method in the physics of these systems, that allow us to
design theoretical experiments tested with a computer. Numerical
simulations define the basic characteristics of the models and
gives a useful path to their theoretical understanding.

From a mathematical point of view the problems explored are
particularly difficult because they consist of iterative systems
with many degrees of freedom and irreversible dynamics. Very
little can be predicted {\it a priori} for systems of this
complexity even when sometimes it is not very difficult to pose a
model that capture the essential features of the phenomena.

While the great majority of the theoretical activity is based upon
toy models it is very important to build a bridge between theory
and real experiments and this is another basic task of computer
simulations. To do that, it is necessary the development of models
more realist and large scale simulations which can be used also in
material characterization. The consequence of this approach is the
application of fractal concepts to the solution of particular
experimental problems (Oil industry, disordered materials, phase
nucleation, crystal growth etc.). The theoretical effort in this
field can be separated into phenomenological or scaling theories
and microscopic theories. At a phenomenological level scaling
theory, inspired to usual critical phenomena, has been
successfully used. This is the nexus to the understanding of the
results of computer simulations and experiments. This method
allows us to identify the relations between different properties
and exponents and to focus on the essential ones. Moreover, it
will be useful to gain more insight into the microscopic dynamics
that evolves into the macroscopic behavior. The connection between
the microscopic and macroscopic behavior has not been extensively
studied.

In this work we make a review of models of growth in quenched
disordered media. In Sec.~\ref{sec:scal} we introduce the dynamics
and static scaling used in experiments and models of growth. We
make also a discussion about the principal source of disorder.
In Sec.~\ref{sec:exp} we make a short summary of the main
experiments on fluid-fluid displacement. In Sec.~\ref{sec:phen} we
present the most used phenomenological equation that is said to
belong to the same universality class than the experiments and/or
models with quenched media in special to the directed percolation
depinning (DPD) models. In Sec.~\ref{TL} we present the Tang and
Leschhorn model \cite{Tang}. In Sec.~\ref{sec:Buld} we present the
Buldyrev {\sl et al.} model \cite{Buldy}. In Sec.~\ref{sec:tl}
some comparisons between both models. In Sec.~\ref{does} we argue
why the Kardar-Parisi-Zhang equation \cite{kpz} with quenched
noise does not describe the DPD models even if they are said to
belong to the same universality class. In Sec.~\ref{sec:ecc} we
deduce an analytical stochastic differential equation for one of
the DPD models starting from its microscopic equation. Finally in
Sec.~\ref{summ} we conclude with a discussion.

\section{Scaling properties of growth in quenched media}
\label{sec:scal}

The investigation of rough surfaces and interfaces has attracted
much attention, for decades, due to its importance in many fields,
such as the motion of liquids in porous media, growth of bacterial
colonies, crystal growth, etc. Much effort has been done in
understanding the processes that induces the roughness in these
fields. When a fluid wet a porous medium a nonequilibrium
self-affine rough interface is generated. The interface has been
characterized through scaling of the interfacial width
$w=\langle[h_i-\langle h_i\rangle]^2\rangle^{1/2}$ with time $t$
and lateral size $L$. The result is the determination of two
exponents $\beta$ and $\alpha$ called dynamical and roughness
exponents respectively. The interfacial width follows
\begin{eqnarray}
w\sim L^\alpha\hspace{1cm}&\mbox{if}&\hspace{1cm} t\gg
L^{\alpha/\beta}\;,\\ w\sim
t^\beta\hspace{1cm}&\mbox{if}&\hspace{1cm}t\ll
L^{\alpha/\beta}\;.\label{scaling}
\end{eqnarray}
The crossover time between this two regimes is of the order of
$L^{\alpha/\beta}$.

The disorder affects the motion of the interface and leads to its
roughness. The main disorder proposed has been the ``annealed''
noise that only depends on time and the ``quenched'' disorder due
to the inhomogeneity of the media where the moving phase is
propagating. While the annealed disorder pushes forward the
interface the quenched disorder brakes this advancement. This last
disorder can be due to impurities, fluctuations in the capillary
size or it could represent all the effects of an inhomogeneous
media.

Some experiments such as the motion of liquids in porous media,
where the disorder is quenched, are well described by the directed
percolation depinning model. Some other experiments like the flow
of fluid trough a disordered media, where the disorder is also
quenched \cite{Horv2,Rubio,He}, give roughness exponents
scattered between $0.6$ and $1.25$ questioning the existence of
universality, the foundation of the scaling hypothesis
(\ref{scaling}).

\section{Experiments on fluid-fluid displacement}
\label{sec:exp}

Several fluid-fluid displacement experiments that can be described
in terms of simple self-affine structures were motivated by the
growing interest of physicians on the dynamics of rough surfaces.
The experiments are easy to perform but the process are very
complex to understand. One of the experiments that was first
performed in order to explain the fact that the roughness exponent
$\alpha$ was not predicted by equations with thermal noise was
carried out by Rubio {\sl et al.} \cite{Rubio} in $1+1$
dimensions. In this experiment the air is displaced by a dyed
water in a porous media consisting of glass beads packed randomly
into a thin horizontal cell of glass covered with
Teflon\footnote{DuPont trademark.}. The water was injected by one
of its edges wetting the porous media and the fluid-fluid
interface is digitized. The roughness exponent was measured over a
distance $\ell$ obtaining a value of $\alpha=0.73 \pm 0.03$. This
value does not depend on the bead diameter and the capillary
number $Ca_{p}$\footnote{$Ca_p \sim a^2 v/\Gamma$, where $a$ is
the diameter of the bead, $v$ is the velocity of the water and
$\Gamma$ is the interfacial tension.} in the range that they
worked. Horv\'ath {\sl et al.} \cite{Horv2} reanalyzed the
interface of Rubio {\sl et al.} and obtained $\alpha=0.81 \pm
0.08$. This discrepancy is not clear, but it could be assign to
the sensibility of this exponent to the details of the data
analysis. Also in this experiment the exponent $\beta \sim 0.65$
was obtained. Other value of $\alpha = 0.63 \pm 0.04$, where the
exponent was measured in the saturation, was obtained by Buldyrev
{\sl et al.} \cite{Buldy,Buldy2} with aqueous fluids absorbed in
many kind of papers. In these experiments, the scaling value does
not depend on the temperature, humidity and kind of paper. They
inferred that the evaporation is irrelevant. Horv\'ath and Stanley
\cite{Horv} performed an experiment in which the evaporation was
controlled by confining the paper between two transparent polymer
sheets. The lower end of the cell was immersed on liquid. The
shape of the interface was digitized and the mean height of the
interface was calculated in real time in order to maintain it at a
fixed height. This prevents any change in the velocity. They found
$\beta=0.56 \pm 0.03$.

\begin{table}[h]
  \centering
\begin{tabular}{|c|c|c|c|}
\hline\hline
  Experiments & $\alpha$ & $\beta$ & References \\ \hline

  Flux of fluid   & $0.73$ & - & \cite{Rubio} \\
                    & $0.81$ & $0.65$ &\cite{Horv2} \\
                    & $0.65-0.91$ & - & \cite{He}\\ \hline
  Wetting paper   & $0.63$  (d=1)  & $0.56$ &\cite{Buldy,Horv}  \\
                    &$0.5$ (d=2)&   -\  &\cite{Buldy2}\\
                    &$0.62-0.78$ (d=1)&0.3-0.4&\cite{Fam3} \\ \hline
  Bacterial growth & $0.78$ & - & \cite{Vic} \\ \hline

  Combustion front & $0.71$ & - &\cite{Zhang1}  \\
\hline\hline
\end{tabular}
\caption{Measured exponents from experiments}\label{Tabla1}
\end{table}
The scattering between the exponents obtained in these experiments
puts in doubt the universality of their results. Sometimes, the
uncertainties could be the result of complex crossover behaviors
before the asymptotic regime is reached. On the other hand,
sometimes, the asymptotic regime is not reached in the
experiments. Also in many experiments the interval where the
exponents were measured is very short enough (less than two
decades), in these cases the results obtained are less reliable.
This does not mean that the idea of scaling behaviors must be
rejected, moreover some scaling hypothesis must be reviewed. In
table \ref{Tabla1} we shows some experimental results.

\section{Phenomenological equation of growth in quenched media}
\label{sec:phen}

Motivated by the success of the Kardar-Parisi-Zang (KPZ)
\cite{kpz} equation in describing the interface motion with
thermal noise, it has been proposed that many interfaces in porous
media are described by the quenched KPZ (QKPZ) equation
\begin{equation}
 \frac{\partial h({\mathbf{x}},t) }{\partial t}=\;{\cal{F}}+\nu \nabla ^2 h +
\frac{\lambda}{2} (\nabla h)^2 + \eta({\mathbf{x}},h) \label{qkpz}
\end{equation}
where $\eta({\mathbf{x}},h)$ represent the quenched white noise in
the media and ${\cal{F}}$ is the driving force responsible of the
advance of the interface. This equation predicts the existence of
a depinning transition: for driving forces ${\cal{F}}
< {\cal{F}}_c$ the interface is pinned, and for ${\cal{F}}
>{\cal{F}}_c$ it moves with a velocity $v \sim f_{red}^{\theta}$
where $\theta$ is the velocity exponent and
$f_{red}=({\cal{F}}-{\cal{F}}_c)/{\cal{F}}_c$ is the reduced
force. Numerical studies \cite{Amaral3,Amaral} indicated that the
discrete models can be grouped into two universality classes
depending on the behavior of the QKPZ nonlinearity $\lambda$.
Isotropic models, {\sl i.e.} models that have no growth direction
determined by the random forces, have $\lambda=0$ or $\lambda \to
0$ as $f_{red} \to 0$. The scaling exponents can be determined
analytically, in one dimension, from Eq.~(\ref{qkpz}) with
$\lambda=0$, obtaining $\alpha_i=1$, $\beta_i=0.75$ and
$\theta_i=0.33$ \cite{Nattermann}. However, anisotropy can induce
a relevant $\lambda$ at the depinning transition. The exponents
characterizing this anisotropic universality can be obtained by an
exacting mapping to directed percolation, obtaining
$\alpha_a=0.63$, $\beta_a=0.63$ and $\theta_a=0.63$
\cite{Tang,Horv}. Many numerical simulations have been done from
a tilted interface \cite{Amaral3,Amaral2} in order to justify if
the nonlinear term of the QKPZ is present in the dynamics of the
proposed models. Recently R\'eka {\it et al.} \cite{Reka}
implemented a method that apply to numerical models and
experimental data and allows to classify the data into one of the
two universality classes without relying on the determination of
the exponents. They assumed that the local velocity has the same
scaling behavior than the global velocity $v$, then
$u({\cal{F}},s) \sim [{\cal{F}}
-{\cal{F}}^{'}_c(s)]^{\theta^{'}}$, where $u({\cal F},s)$ is the
averaged local velocity of the interface with slope $s$,
${\cal{F}}^{'}_c(s)$ is the depinning threshold corresponding to
these segments, and $\theta^{'}$ is the local velocity exponent.
In isotropic media ${\cal{F}}^{'}_c(s)={\cal{F}}_c$ and
$\theta^{'}=\theta$, so the ratio ${\cal{V}} \equiv
u({\cal{F}},s)/v({\cal{F}})$ does not depend on the driven force.
In contrast, in anisotropic media the depinning threshold,
${\cal{F}}^{'}_c(s)$ decreases with $s$ and $\theta^{'}=1$. So, for
anisotropic media $\cal{V}$ has a systematic dependence on
${\cal{F}}$. Plotting ${\cal{V}}$ as a function of $s$ for the
several models and for the fluid-fluid experiment they concluded
that this last experiment belongs to the isotropic class. All
these numerical results confirms the relevance of a nonlinear term
for any model but they do not assert to prove if these models are
well represented by the QKPZ equation or its linear version, the
quenched Edward-Wilkinson (QEW) \cite{ew}.

In particular in models with quenched noise belonging to the
anisotropic class the macroscopic dynamics is influenced by the
coupled effect of the interface itself and the disorder as we
shall see in Sec.~\ref{sec:ecc}.

A powerful method to derive the macroscopic behavior of these
models is to attempt to write the microscopic equation for the
evolution of the local height in these models. This is done {\it
via} the master equation approach \cite{Vvedensky} or directly by
writing the microscopic rules for the evolution of the local
height \cite{Brauns2,Brauns1,Brauns5}. These two methods have
been proved to be equivalent \cite{Costanza}. These microscopic
equations that captures the mechanisms of the growth for each
model must reproduce the macroscopic behavior of the relevant
observable that in this case is the interface itself. Its
continuum version, in a coarsed grained scale is the equation that
represent its universality class. Let us to introduce the two main
models describe the principal features of the experiments of fluid
imbibition on paper sheet \cite{Buldy}.

\section{Growth mechanisms for the Tang and Leschhorn model}
\label{TL}

We present now the microscopic Tang and Leschhorn (TL) model. The
interface growth takes place in a lattice of $N$ points of edge
$L$ ($N=L/a$) where $a$ is the lattice constant. Usually in discrete
models $a$ is taken as one, so $N=L$. We will retain the lattice
constant because we will use it below.

Periodic boundary conditions are used. The quenched noise is
represented by a random pinning force $g(\mathbf{r})$ uniformly
distributed in $[0,1]$ assigned to every cell of the lattice. This
random force models the random distribution of the fiber of paper.
For a given pressure $p$, that represents the driving force that
push the fluid into the media, the cells are divided in two
groups, active (free) cells with $g(\mathbf{r})\le p$ and inactive
(blocked) cells with $g(\mathbf{r})>p$. Notice that $q=1-p$ is the
density of blocked cells. It is well known that there
exist a critical density of blocked cells $q_c=0.539$, what is the
directed percolation threshold for an infinite lattice, above $q_c$
the interface becomes pinned.

In this model, the growth event is defined as follow:
\begin{enumerate}
\item If $h_i$ is greater than either $h_{i-1}$ or $h_{i+1}$ by
one or more units, the height of the lower of the two columns
$(i-1)$ and $(i+1)$ is incremented in $a$ (in case of tie, one of
the two is chosen with equal probability).
\item In the opposite case, $h_i
< \min (h_{i-1},h_{i+1}) +2 a$, the column $i$ advances by one unit
provided that the cell to be occupied is an active cell.
\item Otherwise no growth takes place.
\end{enumerate}
In this model, the time unit is defined as one growth attempt. In
numerical simulations at each growth attempt the time $t$ is
increased by $\delta t$, where $\delta t = 1 / N$.

Notice that the condition for the interface becomes pinned is that:
\begin{itemize}
\item $\bigtriangleup h=\pm a,0$ and
\item all the sites above the interface are blocked.
\end{itemize}
So, the interface becomes pinned if a cluster of inactive sites
connected horizontally or diagonally expands all the lattice. This
happens for $q\geq q_c$. This define a cluster of directed
percolation. For $q < q_c$ the interface is characterized by two
correlations lengths that behave approaching to the critical
point as,
\begin{eqnarray}
\xi_\parallel &\sim& |q-q_c|^{-\nu_\parallel}\label{xipar}\;,\\
\xi_\perp &\sim&|q-q_c|^{-\nu_\perp}\;,
\end{eqnarray}
which are the characteristic lengths of these clusters in both
directions, with $\nu_\parallel=1.733$ and $\nu_\perp=1.097$.

The interface is specified by a set of integer column heights
$h_i$ ($i=1,\dots,N$). During the growth, a column is selected at
random with probability $1/N$ and compared its height with those
of its neighbors. In a temporal step the height in the site $i$ is
increased by,
\begin{enumerate}
\item\hspace{.5cm}
$a$ \hspace{1.4cm} if $j=i+1$ and $h_{i+1} \ge h_i + 2 a$ and $h_i
< h_{i+2}$,
\item\hspace{.5cm}
$a/2$ \hspace{1cm} if $j=i+1$ and $h_{i+1} \ge h_i + 2 a$ and $h_i
= h_{i+2}$,
\item\hspace{.5cm}
$a$ \hspace{1.4cm} if $j=i-1$ and $h_{i-1} \ge h_i + 2 a$ and $h_i
< h_{i-2}$,
\item\hspace{.5cm}
$a/2$ \hspace{1cm} if $j=i-1$ and $h_{i-1} \ge h_i + 2 a$ and $h_i
= h_{i-2}$,
\item\hspace{.5cm}
$a$ \hspace{1.4cm} if $j=i$ and $h_i < \min(h_{i-1},h_{i+1}) + 2
a$ and $G_i(h_i+a)=1$.
\end{enumerate}
Notice that these are the rules for the growth in this column. In
this context this way to describe the evolution equation is
equivalent to derive the first moment from the master equation. In
Fig.~\ref{esqtl} we show this five contributions to the growth of
the site $i$.
\begin{figure}
\vspace{65mm}
\caption{Schematic representation of the 5 cases that contributes
to the growth of the $i$-th site. The arrows show the elected
site. In grey it is shown the columns wetted at the time $t$. In
all the cases the $i$-th column will grow in one unit at the time
$t +\delta t$ .\label{esqtl}}
\end{figure}

The evolution equation \cite{Brauns2} for the interface in a time
step $\delta t=1/N$ is
\begin{equation}
h_i(t+\delta t) = h_i(t)+\delta t\, a\,R_i\;,\label{me}
\end{equation}
with
\begin{equation}
R_i=W_{i+1}+W_{i-1}+G_i(h'_i)\,W_i\;,\label{Ri}
\end{equation}
and
\begin{eqnarray}
W_{i\pm 1}&=&\Theta(h_{i\pm 1}-h_i-2 a)
    \{[1-\Theta(h_i-h_{i\pm 2})]+\delta_{h_i,h_{i\pm 2}}/2\}\;,\\
W_{i}&=&1-\Theta(h_i-\min(h_{i-1},h_{i+1})-2 a)\;.\label{wi}
\end{eqnarray}
Here $h'_i=h_i+a$ and $\Theta(x)$ is the unit step function
defined as $\Theta(x)=1$ for $x\ge 0$ and equals to $0$ otherwise.
$G_i(h_i')$ equals to $1$ if the cell at the height $h_i'$ is free
or active ({\sl i.e.} the growth may occur at the next step) or
$0$ if the cell is blocked or inactive. $G_i$ is called the
interface activity function. Taking the limit $\delta t \to 0$ and
averaging over the lattice we obtain ($h=\langle h_i\rangle$)
\begin{equation}
\frac{dh}{dt}= \langle 1-W_i \rangle + \langle G_i
W_i\rangle\label{dhdt}\;.
\end{equation}
Here we used $a=1$ as is usual in a discrete model. This equation
allow us the identification of two separate contributions: the
lateral $\langle 1-W_i\rangle$ and the local one $\langle
G_i\,W_i\rangle$.

The temporal derivative of the square interface width (DSIW) is:
\begin{equation}
\frac {d w^2}{dt} = 2\, \langle(h_i-\langle
h_i\rangle)\,R_i\rangle\;.\label{dwdt}
\end{equation}
The DSIW can also be expressed by means of local and lateral
additive contributions. The lateral contribution is
\begin{equation}
2\,[\,\langle(1-W_i)\,\min(h_{i-1},h_{i+1})\rangle- \langle
1-W_i\rangle\,\langle h_i\rangle\,]\label{dsiw_d}\;,
\end{equation}
and the local contribution is
\begin{equation}
2\,[\,\langle h_i\,G_i\,W_i\rangle - \langle h_i\rangle\,\langle
G_i\,W_i\rangle\,]\label{dsiw_s}\;.
\end{equation}
In order to obtain the lateral contribution [Eq.~(\ref{dsiw_d})]
we used that $\Theta(x-y)+\Theta(y-x)-\delta_{x,y}=1$.

\begin{figure}
\vspace{150mm} \caption{DSIW (solid line), and its lateral
($\bigcirc$) and local ($\Box$) contributions vs $\ln t$; for $q$
equal to $0.3$ (A), $0.539$ (B) and $0.6$ (C).} \label{DW}
\end{figure}

In Figure~\ref{DW} we plot both contributions as a function of time
for various values of q. At short times, the lateral process is
unimportant because $\Delta h$ is mostly less than two. As $t$
increases, the behavior of this contribution depends on $q$.
Notice, from Eq.~(\ref{dsiw_d}), that the lateral contribution may
be either negative or positive. The negative contribution tends to
smooth out the surface. Figure~\ref{DW} shows that this case
dominate for small $q$. The positive lateral contribution enhances
the roughness. This last effect is very important at the critical
value. At this value, the local contribution is practically
constant, but the lateral contribution is very strong, enhancing
the roughness. This last contribution has important duties on the
power law behavior. Generally speaking, the local contribution
roughen the interface while the lateral one flatten it for small $q$,
but the lateral contribution also roughen the interface when $q$
increases. The lateral contribution is enhanced by local growth.
The lateral growth  may also increase the probability of local
growth. This crossing interaction mechanism makes the lateral
growth dominant near the criticality.

\section{Growth mechanisms for the Buldyrev {\sl et al.} model}
\label{sec:Buld}

The interface growth takes place under the same initial conditions
as the TL model. During the growth, a column is selected at random
with probability $1/L$ and the highest dry active cell, in the
chosen column, that is nearest-neighbor to a wet cell is wetted.
Afterwards, we wet all the dry cell below it. The height is
increased by \cite{Brauns5}:
\begin{enumerate}
\item\hspace{.5cm}
$a$ \hspace{1.5cm} if $h_i \ge \max(h_{i+1},h_{i-1})$
\hspace{.5cm}and\hspace{.5cm} $G_i(h_i+a)=1$,
\item\hspace{.5cm}
$a\,Y_i$ \hspace{1cm} if $h_i < \max(h_{i+1},h_{i-1})$
\hspace{.5cm}and\hspace{.5cm} $G_i(h_i+a\,Y_i)=1$,
\end{enumerate}
where
\begin{equation}
 Y_i=\sum_{k=1}^{z_i} k G_i(h_i+k a) \prod_{j=k+1}^{z_i}
\left(1-G_i(h_i+ j a)\right)\label{yi}
\end{equation}
and
\begin{equation}
a z_i=\max (h_{i-1},h_{i+1}) - h_i\;,
\end{equation}
\begin{figure}
\vspace{70mm}
\caption{Schematic representation of the growth
mechanism in the $i$-th site. The region in grey shows the height
at time $t$, in light grey it is shown the height at the time $t
+\delta t$. \label{esqbuld}}
\end{figure}
with $G_i(h_i+j)$ equal to 1 if the cell is active and 0 if the
cell is inactive. In Fig.~\ref{esqbuld} we represent schematically
the rules. Notice that in this model the condition for the system
get pinned is that a cluster of blocked sites, connected by
nearest-neighbor expand all the lattice. This is achieved, in the
static regime for $q \geq q_c$ with $q_c=0.469$. In this model,
the time unit is defined as one growth attempt. In numerical
simulations, at each growth attempt, the time $t$ is increased by
$\delta t=1/N$. In this way, after $N$ growth attempts the time is
increased in one unit. We consider the evolution for the height of
the $i$-th site of the process described above. Let us denote by
$h_i(t)$ the height of the $i$-th generic site at time $t$.
Freezing the simulation at a given time, we compute the temporal
evolution for the interface height in the next time as
\begin{equation}
h_i(t+\delta t) = h_i(t) + \delta t\;a \{
\Theta(-z_i)\,G_i(h_i+1)+[1-\Theta(-z_i)]\,Y_i\}\label{first}\;,
\end{equation}
where $\Theta(x)$ was defined in Sec.\ref{TL}. $Y_i$ is the
increase of the height in the $i$-th column due to the
contribution of the nearest-lateral-neighbor. Notice that, this
kind of growth occurs by wetting the active cells of the chosen
site nearest-neighbor of a wet cell by lateral contact, and then
eroding all the cells bellow from a wet cell. We shall call
contact contribution to the term $[1-\Theta(-z_i)]\,Y_i$ and local
contribution to the term $\Theta(-z_i)\,G_i(h_i+1)$ from
Eq.~(\ref{first}).

Averaging over the lattice, taking $\delta t\to 0$, the evolution
equation for the square interface width is
\begin{equation}
\frac{d w^2}{d t}= 2 \langle (h_i-\langle h_i\rangle)
\Theta(-z_i)\,G_i\rangle + 2 \langle (h_i-\langle h_i\rangle)
(1-\Theta(-z_i))\,Y_i\rangle\;.
\end{equation}
The first term of both equations can be identified as the local
growth contribution, and the second term as the contact growth
contribution. In the present work we focus only on the dynamical
behavior for the roughness.
\begin{figure}
\vspace{65mm} \caption{DSIW as a function of time. The parameter $p$
is 0.7 ($\bigcirc$), 0.56 ($\bigtriangledown$), 0.531 ($\bullet$)
and 0.51 ($\bigtriangleup$). The symbol $\bullet$ shows the
critical behavior.} \label{dw2dt}
\end{figure}
Figure \ref{dw2dt} shows the temporal derivative of the square
interface width (DSIW) as a function of time for various values of
$p$. The initial condition is $p$ in all regimes. As we expected
\cite{Brauns3}, the power law holds only at the criticality. The
DSIW goes asymptotically to zero at the pinning and moving phase.
In Figure \ref{dw2dt2}, we show the two contributions to the DSIW
for different values of $p$. The local contribution $2 \langle
(h_i-\langle h_i\rangle)\,\Theta(-z_i) \,G_i\rangle$ to the DSIW
is always positive. As $p$ decreases this contribution becomes
less important, but always rough the interface. On the other hand,
for $p>p_c$, the contact contribution $2 \langle (h_i-\langle
h_i\rangle)\,(1-\Theta(-z_i))\,Y_i\rangle$ can take negative
values, smoothing out the surface. Otherwise, for $p\le p_c$, the
contact contribution is always positive roughening the interface.
One could expect that the contact contribution always smooth out
the surface because it tends to widen the roughen picks. However,
near the criticality, the contact growth happens mainly in lateral
neighbors cells to few height terraces above the mean height.
Then, this new wetted column smooth out locally, but it moves away
from the mean height increasing the roughness.
\begin{figure}
\vspace{150mm} \caption{Semi-ln plots of the different
contributions to the DSIW as a function of time for the Buldyrev
{\sl et al.} model for different values of $p$. The circles
represent the contact contribution and the squares represent the
local contribution for the moving, the pinning and the critical
regime, respectively.} \label{dw2dt2}
\end{figure}

\section{Comparisons between the Buldyrev {\sl et al.} and the Tang and
Leschhorn models} \label{sec:tl}

We rescue the similarities between the Buldyrev {\sl et al.} and
the Tang and Leschhorn models. In spite of the strong microscopic
differences between their rules, the results obtained from
the microscopic equation are similar. The main conclusion is that
in both models the lateral (contact) term plays an important role
in the power law behavior. The origin of the behavior of the
lateral term near the criticality arises from a nonlinear term.
This could lead to think that the QKPZ equation contains these
features, but this is not the case as we shall show (see Sec.
\ref{sec:ecc}). Moreover, the QKPZ does not allow to explain the
cross mechanism between the two contribution, as we shall see
below.

\section{Does the QKPZ describe the DPD models?}
\label{does}

Let us explain first why the QKPZ equation does not take into
account this coupled effect. In the QKPZ we can distinguish two
contributions, the local growth ${\cal{S}}= {\cal{F}} +\eta(x,h)$
and the lateral one ${\cal{L}}=\nu\;\partial^2_x h +
\frac{\lambda}{2}\; (\partial_x h)^2$. So, we can write the
evolution equation for the height $h=h(x,t)$ as \cite{Brauns6}
\begin{equation}
\partial_t h={\cal{S}} +{\cal{L}}\;.
\end{equation}
Taking the derivative of the square interface width, $w^2=\langle
(h-\langle h \rangle)^2\rangle$, its evolution equation is given
by
\begin{equation}
{\partial_t w}^2=2\langle (h-\langle h \rangle){\partial_t
h}\rangle=
 2\langle (h-\langle h \rangle){\cal{S}}\rangle+
2\langle (h-\langle h \rangle){\cal{L}}\rangle
\end{equation}

The first term can be identified as the local growth contribution,
and the second term as the lateral growth contribution. The
separation into these two analytical terms allows us to compare
the mechanisms of growth in this model with the mechanisms in the
DPD models. We have performed the direct numerical integration of
Eq.~(\ref{qkpz}) in one dimension in the discretized version
\cite{Csahok,Jeong}
\begin{eqnarray*}
h(x,t+\triangle t)&=&h(x,t)+ \triangle
t\;\left\{h(x-1,t)+h(x+1,t)\right.\\&
&-2h(x,t)+\frac{\lambda}{8}\;\left\{h(x+1,t)-h(x-1,t)\right\}^2 \\
& & +{\cal{F}}+\eta(x,[h(x,t)])\left. \right\}
\end{eqnarray*}
\begin{figure}
\vspace{65mm} \caption{DSIW as a function of time in the critical,
pinning, and moving phases for $\lambda=1$. The parameter
${\cal{F}}$ is 0.464 (solid line), 0.43 (dashed line), and 0.54
(dotted line).} \label{dw2dtT}
\end{figure}
where $[\dots]$ denotes the integer part and $\eta$ is uniformly
distributed in $[-\frac{a}{2},\frac{a}{2}]$, where $a=10^{2/3}$ is
selected. We choose $\triangle t =0.01$, use the initial condition
$h(x,0)=0$ and the periodic boundary conditions. Figure
\ref{dw2dtT} shows the DSIW as a function of time for various
values of ${\cal{F}}$. Here we found that the DSIW increases
continuously from zero to a maximum value, at difference of the
DPD models where $p$ is the initial condition in all regimes.
Moreover we did no expected to recover the initial regime because
the QKPZ equation is valid only in the hydrodynamic limit. Equally
to the DPD models \cite{Brauns2,Brauns3,Brauns4}, the power law
holds only at the criticality. The DSIW goes asymptotically to
zero at the pinning and moving phase. In the asymptotic regime,
the behavior of the DSIW is similar in the QKPZ equation and in
the DPD models.

In Figure \ref{dw2dt2T}, we show the two contributions to the DSIW
for different values of ${\cal{F}}$. The local contribution
$2\langle (h-\langle h \rangle){\cal{S}}\rangle$ to the DSIW is
always positive. As ${\cal{F}}$ decreases, this contribution
becomes less important but always rough the interface. On the
other hand, the lateral contribution $2\langle (h-\langle h
\rangle){\cal{L}}\rangle$ takes negative values in all phases
smoothing out the surface. This is an important difference with
the DPD models where, for $p\leq p_c$, the lateral contribution is
always positive roughening the interface. So, the QKPZ does not
represent exactly the dynamics of the TL model even if the
exponents derived from its numerical integration are in accord
with the ones obtained for these models. The cross mechanism
between both contributions is not taken into account in this
equation because the noise is additive. This cross mechanism is
due to the fact that the quenched noise is coupled to the dynamic
of the interface as was shown by Braunstein {\sl et al.}
\cite{Brauns2,Brauns1,Brauns5}. In order to explain the origin of
this coupling let us introduce the passage to the continuum of our
microscopic equation.
\begin{figure}
\vspace{150mm} \caption{Semi-ln plots of the different
contributions to the DSIW as a function of time for different
values of ${\cal{F}}$ and $\lambda=1$. The circles ($\bigcirc$)
represent the local contribution, the triangles ($\bigtriangleup$)
represent the lateral contribution, and the squares ($\Box$)
represent the total DSIW. The (a) plot shows the critical phase
${\cal{F}}=0.464$. The (b) plot shows the pinning phase
${\cal{F}}=0.43$. The (c) plot shows the moving phase
${\cal{F}}=0.54$.}
 \label{dw2dt2T}
\end{figure}

\section{Stochastic Differential Equation for the TL model}
\label{sec:ecc}

Expanding Eq.~(\ref{me}) to first order in $\delta t$, the
evolution for the height of this site is \cite{Brauns7}
\begin{equation}
\frac{\partial h_i}{\partial t} = \frac{a}{\tau} R_{i} +
\eta_i\label{1}\;,
\end{equation}
where $\tau=N\;\delta t$ is the mean lapse between successive
election of any site and $\eta_i$ is a Gaussian ``thermal'' noise
with zero mean  and covariance $\langle \eta_i(t)\eta_j(t')\rangle
= (a^2/\tau)\; R_i\;\delta_{ij}\;\delta(t-t')$ \cite{Vvedensky}.
For this model,
\begin{eqnarray}
R_{i}(h_{i-1},h_i,h_{i+1})=W_{i+1}+
W_{i-1}+G_i(h_i+a)\;W_i\label{Gi}\;,
\end{eqnarray}
and $W_{i+1}$, $W_{i-1}$ and $W_i$ where defined in
Eq.~(\ref{wi}).

In order to obtain a stochastic equation we need an analytic
representation of $R_{i}$. To do this we proceed to regularize the
height defining an interpolating function \cite{Vvedensky} for the
difference of height and then to expand the step function to first
order in the argument as $ \Theta(x) \approx c_0 + c_1 x
+{\cal{O}}(x^2)$. This can be done providing that $x$ is smooth
obtaining
\begin{equation}
\frac{\partial h(x_i,t)}{\partial
t}=\frac{a}{\tau}\;\left[W(x_i+a)+
W(x_i-a)+W(x_i)\;F(x_i,h(x_i)+a)\right]
+\eta(x_i,t)\label{dhdt1}\;,
\end{equation}
with
\begin{eqnarray}
W(x+a)+W(x-a)\!\!&=&\!\!(c_0- 2c_1) +4\,c_1^2(\partial_x h)^2+a\,
c_1\left[\frac{1}{2}+4\,(c_0-2c_1)\right]\;
\partial_x^{2}h\;,\nonumber\\&&\label{difu}
\\ W(x)\!\!&=&\!\!1-(c_0-2 c_1) - 4\,c_1^2 \;(\partial_x h)^2 +
\frac{1}{2}\,a\,c_1 \;\partial_x^2 h\label{sustrato}\;.
\end{eqnarray}
Notice that the argument of $G=\Theta(p-g(x_i,h(x_i)+a))$ is not
smooth, so its expansion is meaningless. In order to recover the
early time regime where the dynamics is mainly random deposition
with probability $p$ \cite{Lopez,Brauns4} we must impose the
condition $c_0=2\,c_1$. The final step is a coarse-grained spatial
average of the variables in order to obtain smooth continuous
functions at a macroscopic level. In this way, we obtain the
stochastic continuous equation for this model,
\begin{equation}
\frac{\partial h}{\partial t}=\mu(G)+ \nu(G)\;\partial^2_x h +
\lambda(G)\; (\partial_x h)^2 +\eta(x,t)\;,\label{bbac}
\end{equation}
where ($G \equiv G(x,h)$)
\begin{eqnarray*}
\mu(G)&=&G\,\frac{a}{\tau}\;,\\
\nu(G)&=&\frac{1}{2}\,c_1\,(1+G)\,\frac{a^2}{\tau}\;,\\
\lambda(G)&=&4\,c_1^2\,(1-G)\,\frac{a}{\tau}\;.
\end{eqnarray*}

Equation~(\ref{bbac}) shows that the nonlinearity arises naturally
as a consequence of the microscopic model. As we approach to the
critical value, in the dynamical regime, the density of active
sites $f=\langle G \rangle$ goes asymptotically to zero
\cite{Brauns1} and the coefficient of the nonlinear term becomes
relevant. In these case the main responsible of the nonlinearities
is the lateral contribution [see Eq.~(\ref{difu})]. This explain
why this contribution enhances the roughness at the criticality as
was predicted by Braunstein {\sl et al.} \cite{Brauns1}. Faraway
above the criticality the nonlinear term becomes less relevant. In
the limit $p \to 0$ ($p \to 1$) we recover the KPZ (EW)  equation
with thermal noise as was expected. In the asymptotic regime ($t
\gg t^*$) the various derivatives of the height become very small
on a coarse grained scale. In this temporal regime the mean height
speed (MHS) $\langle\partial h /\partial t \rangle \to f$. For $p
\leq p_c$, $f \to 0$ and the MHS goes to zero while for $p \gg p_c$,
$f\to \mbox{const}$ and MHS goes to constant.

Notice that our equation is invariant under local tilting of the
interface by an infinitesimal angle in the same way that the QKPZ
equation (\ref{qkpz}) is \cite{kpz}, so the previous numerical
results obtained by Amaral {\sl et al.} \cite{Amaral}, that
studied the effects of the effective coefficient $\lambda_{eff}$
from a tilted interface, are compatible with our equation. Our
result is also in agreement with those of R\'eka {\sl et al.}
\cite{Reka} that obtained numerically a parabolic shape of the
local velocity as a function of the gradient for the DPD model near
above the criticality for different reduced forces $(p/p_c-1)$.

In the experiments, the advancement of the interface is
determinated  by the coupled effect of the random distribution of
the capillary sizes, the surface tension and the local properties
of the flow, so it is not surprising that all these effect give
rise to a multiplicative noise. This multiplicative noise must be
taken into account at the time to pose a model with the essential
features of the experiment of surface growth in disordered media.
In the TL and the Buldyrev {\sl et al.} models the growing rules
for the evolution of the local height are strongly coupled to the
quenched noise in a multiplicative way. In both models the
microscopic rules that allows the growth from an unblocked cell
\cite{Brauns2,Brauns3} depends in some way on the local slope. In
that sense this coupled effect is not taken into account in the
QKPZ equation. The effect of a multiplicative noise has been
proposed by Csah\'ok {\sl et al.} \cite{Csahok} by means of a
phenomenological equation. They found a crossover between two
temporal regimes with $\beta=0.65$ to $\beta=0.26$ but the value
of $\alpha \simeq 0.47$ was obtained over a short range spatial
scale. Indeed, the exponents are not the same as of the DPD
models. Moreover, processes with the same exponent may not belong
to the same universality class. For example, $1+1$-dimensional
lattice gas simulations of roughening of immiscible fluid-fluid
interface \cite{Flekkoy} lead to the same exponents as the
$1+1$-dimensional KPZ \cite{kpz} ($\beta=1/3$ and $\alpha=1/2$)
for surface growth, but this model is completely linear, so there
is no obvious mathematical relationship between these two
processes.

\section{Summary}
\label{summ}

The microscopic equations for the models with quenched noise,
treated in this work, allows to gain a more profound insight on
the principal mechanism of growth. In special, the separation into
the lateral and the local growth contributions allows to explain
the great interplay between them. Obviously, the continuous
equation that represents the local growth of these processes must
take into account both mechanisms in addition of the symmetries
allowed by the model, but this is not enough to reproduce the
interplay between both mechanisms. The QKPZ equation contains a
lateral and a local contribution but the quenched noise is
additive. Notice that a relevant common feature of the two models
treated in this work is that the rules for growth in active sites
depend strongly on the slope. So, it is not surprising that this
lead to a coupled effect of the quenched noise to the dynamics.
This coupled effect is not taken into account in any equation with
additive noise. Nevertheless, the QKPZ equation gives the same
scaling exponents that these models, moreover it is not clear why.
Our results suggest that the QKPZ equation does not describe
properly the dynamics of the DPD models even if the exponents are
similar. Our Langevin equation for the TL model reflects this
coupled effect through its coefficient noise dependence. The
equation obtained has the same terms than the QKPZ but its
coefficient depends on the competition between the driving force
and the quenched noise. In that sense, our equation is
multiplicative in the noise.

Finally, we conclude that the classification of these models in
universality classes is not fully developed. The derivation of the
continuous equation from the microscopic dynamics is a powerful
method that allows to associate them in a non ambiguous way.

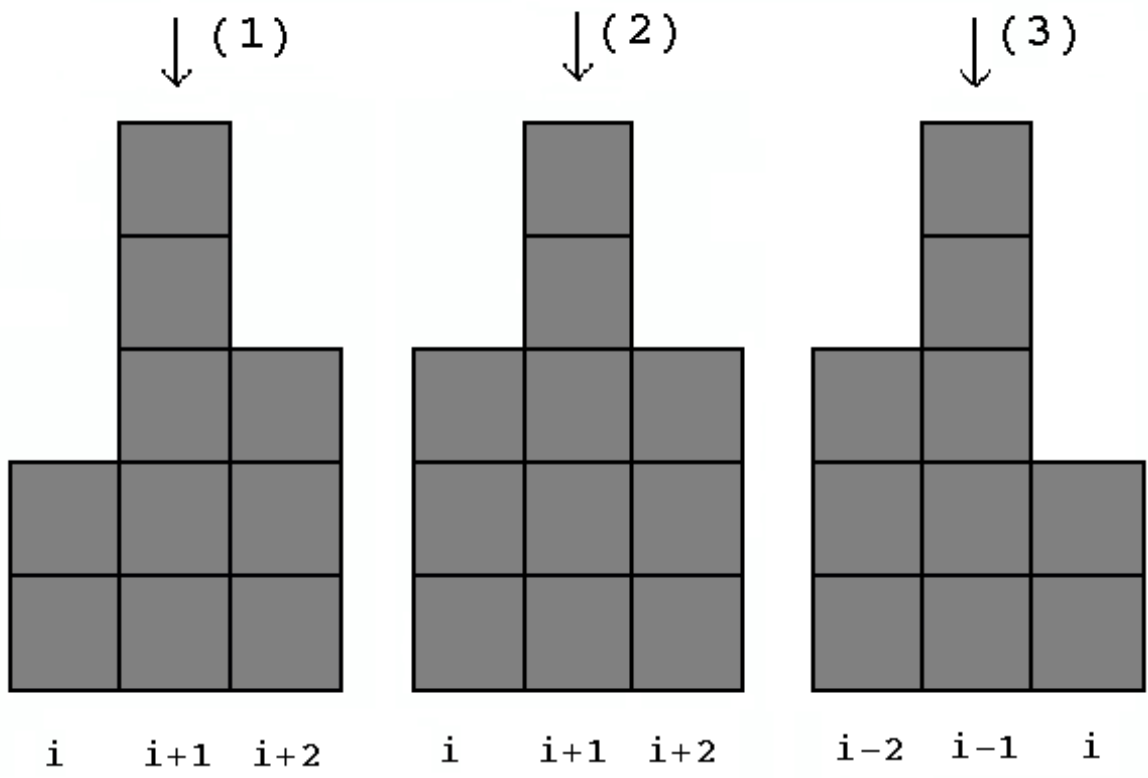
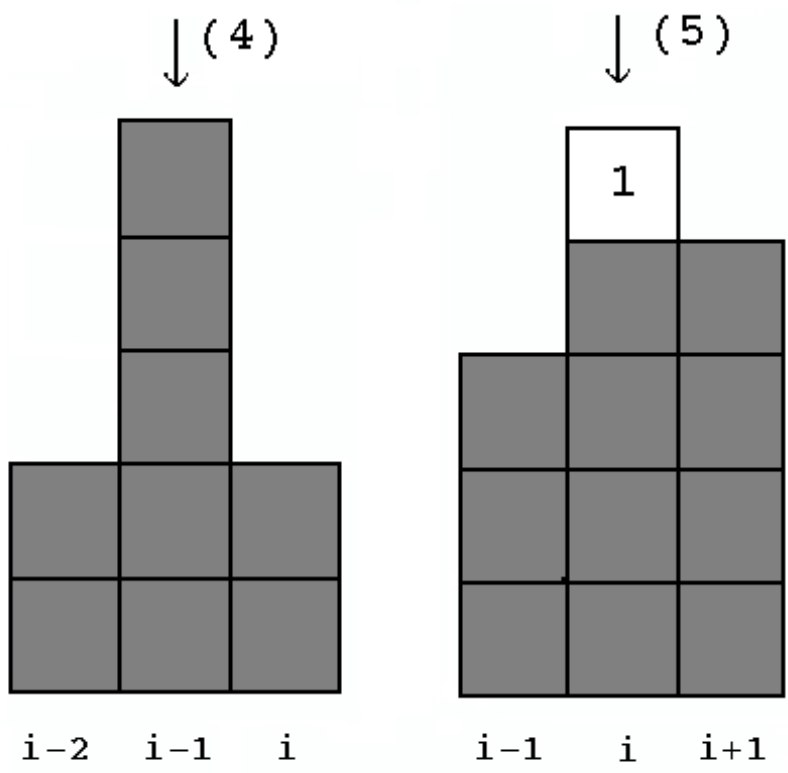
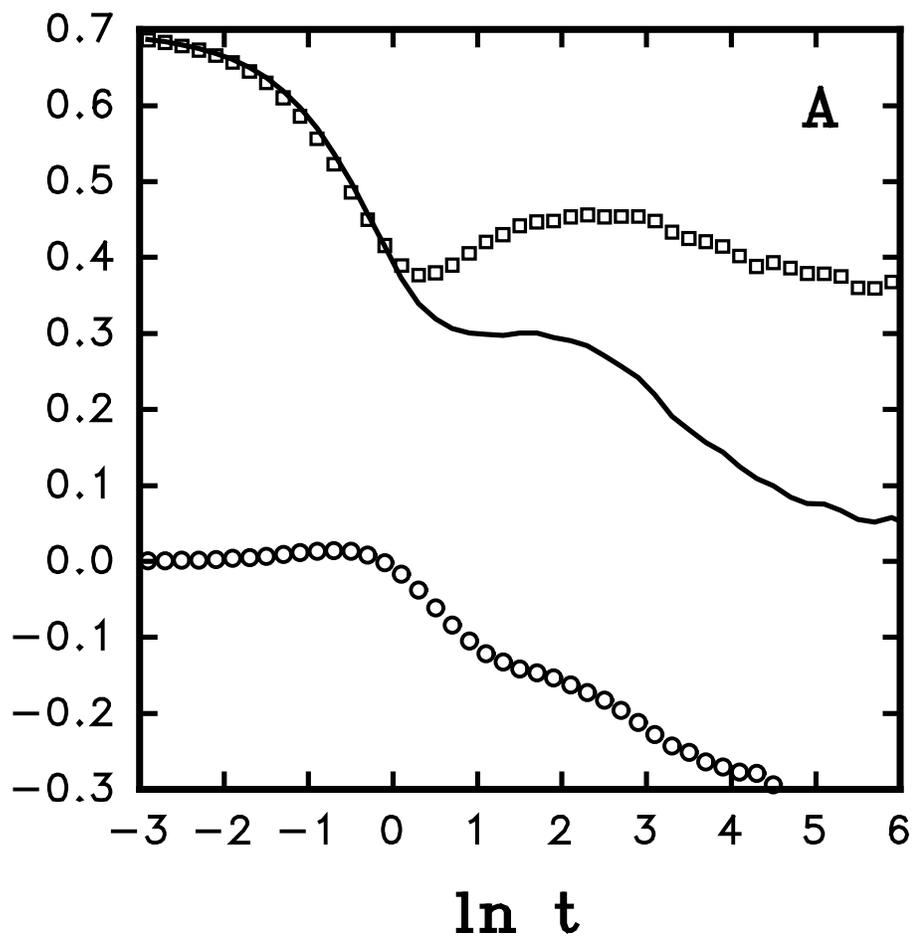
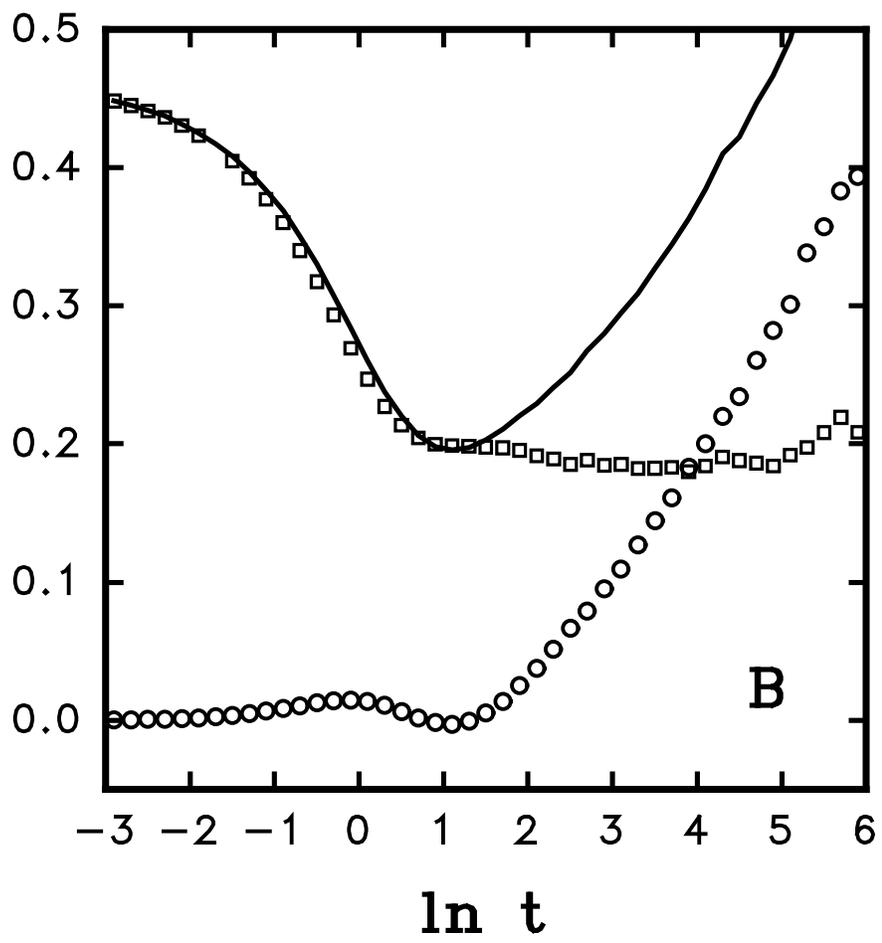
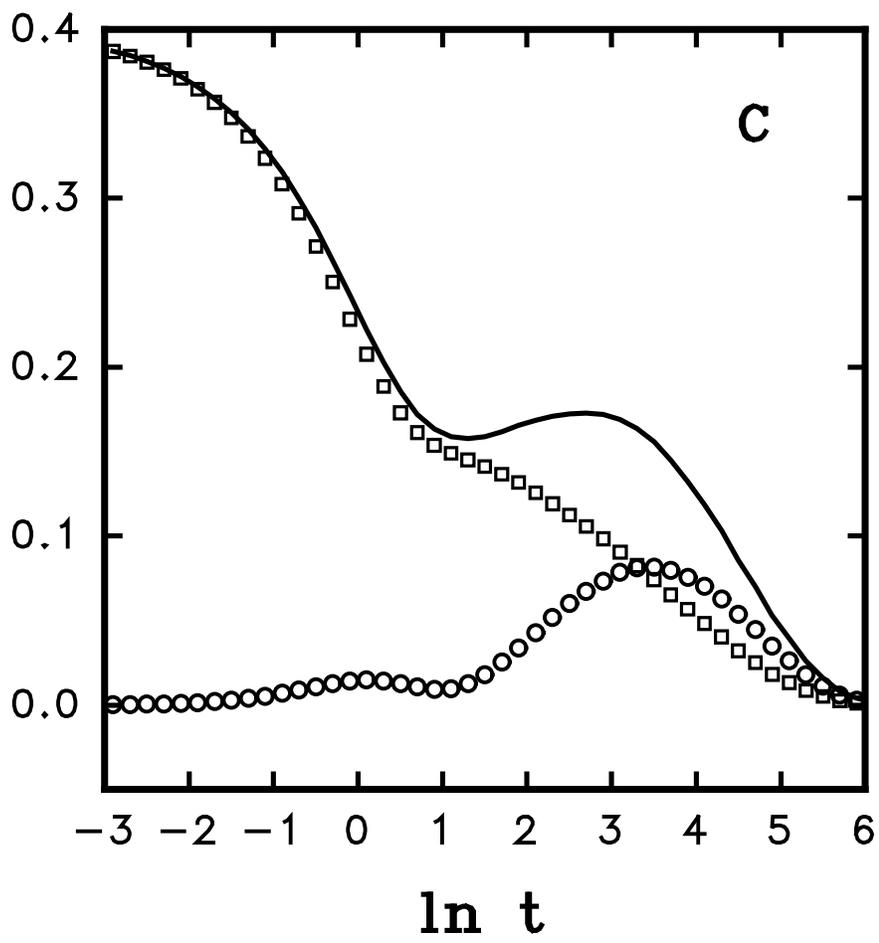
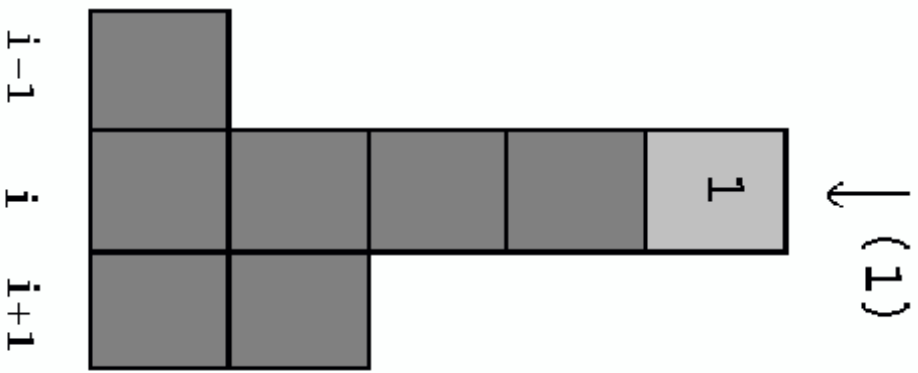
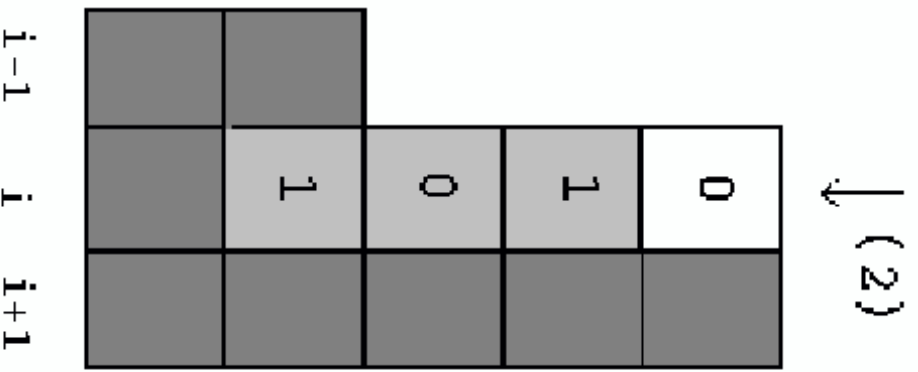
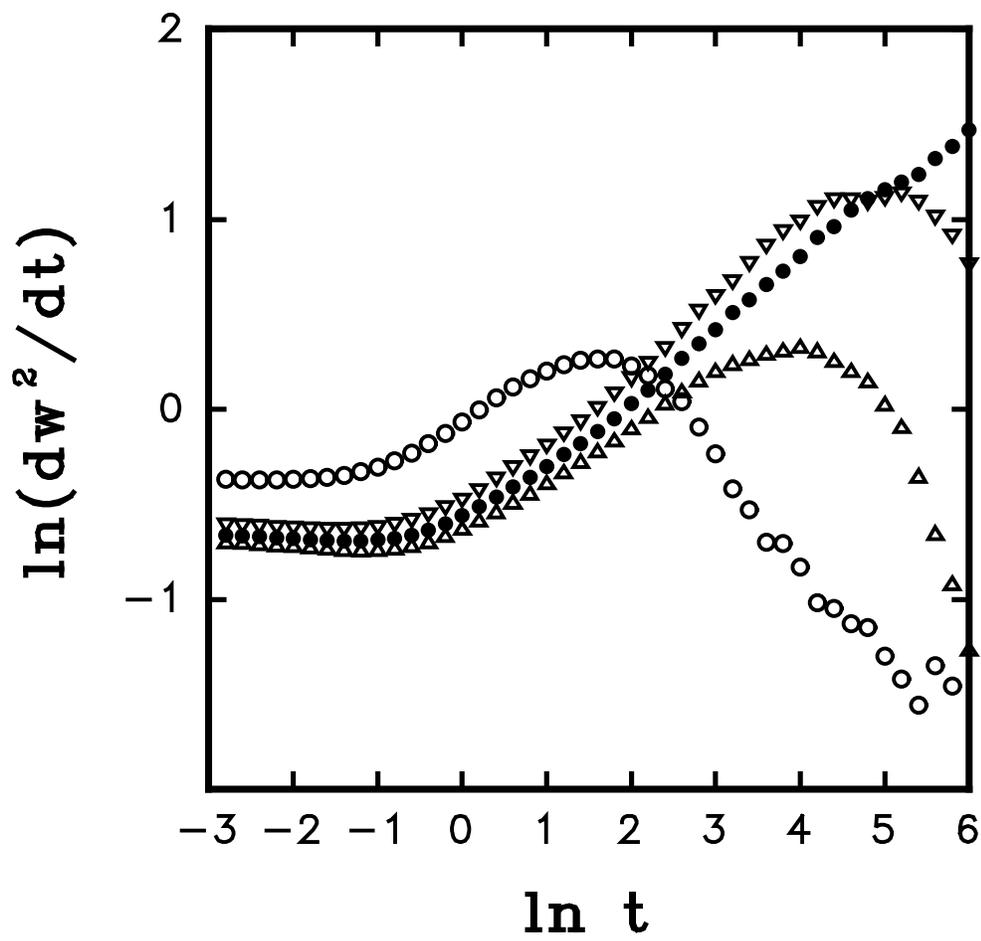
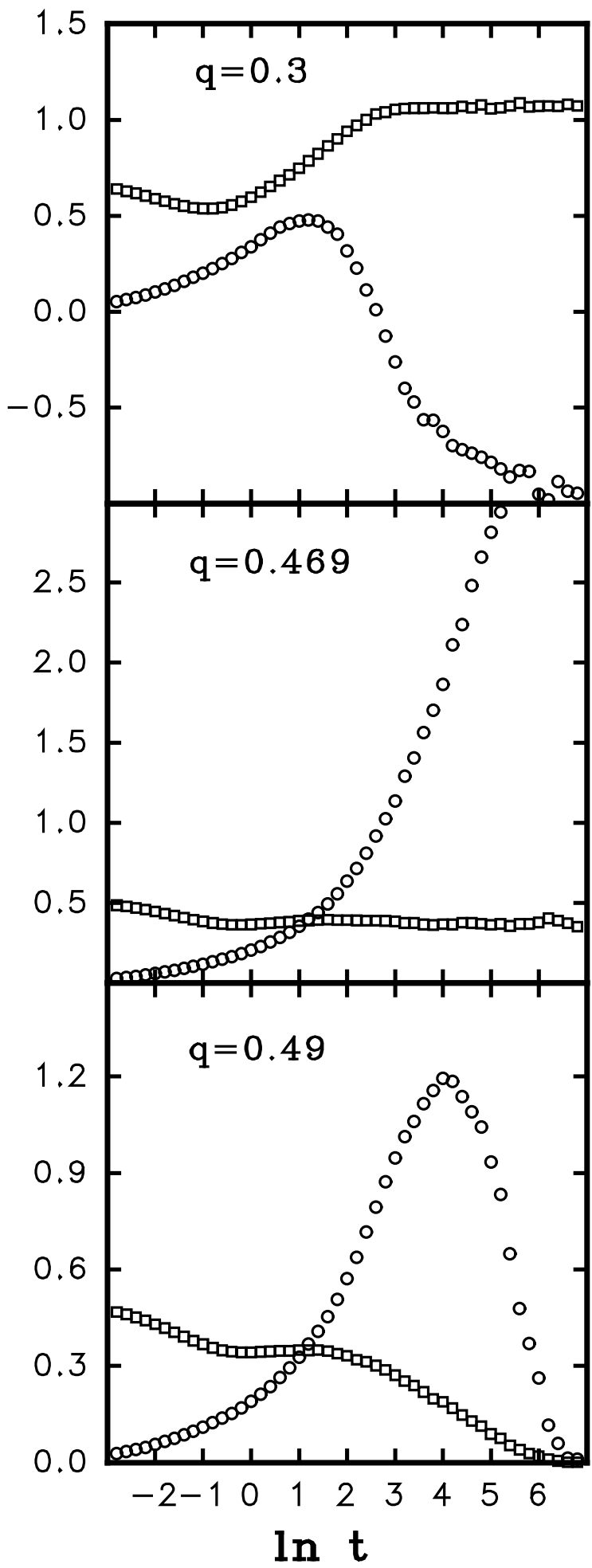
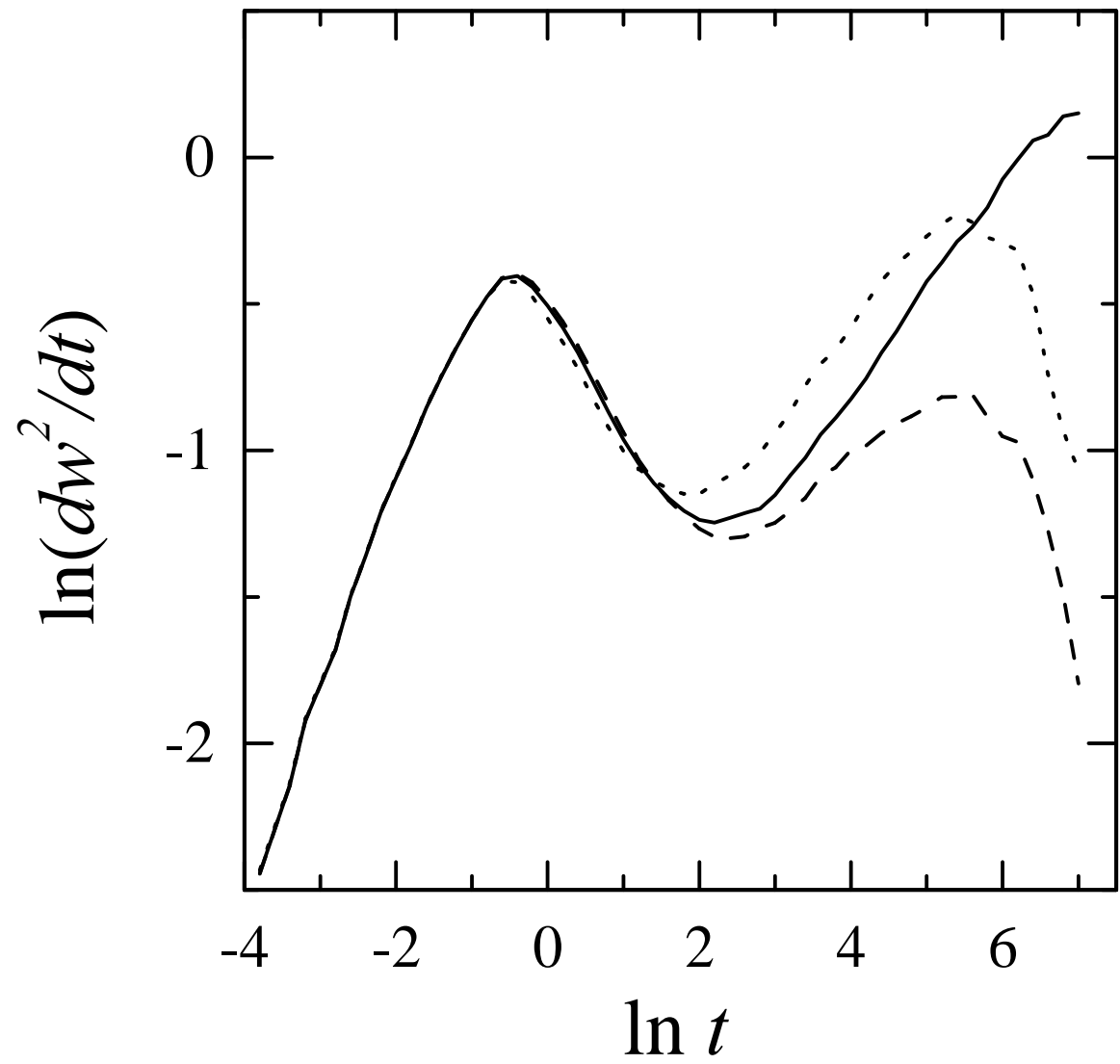
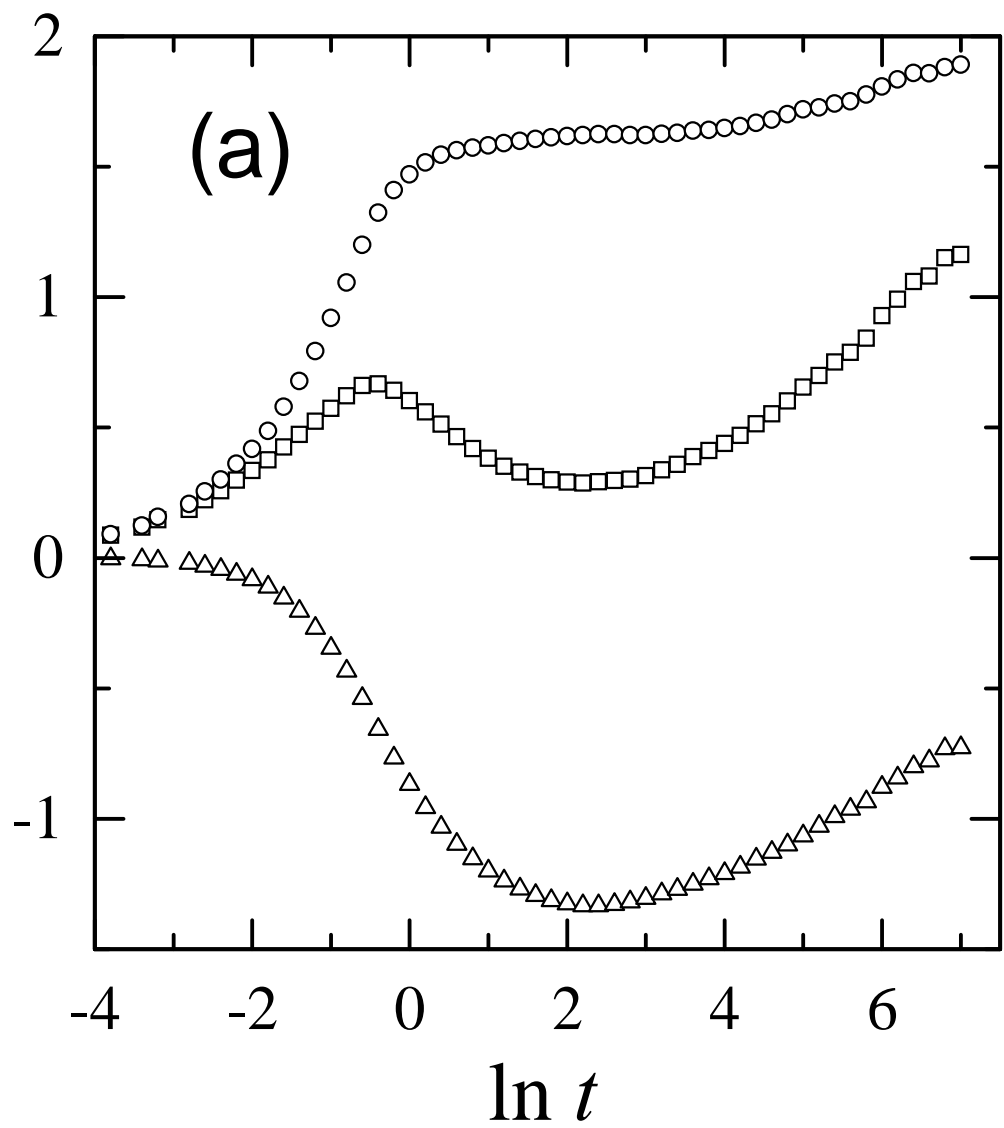
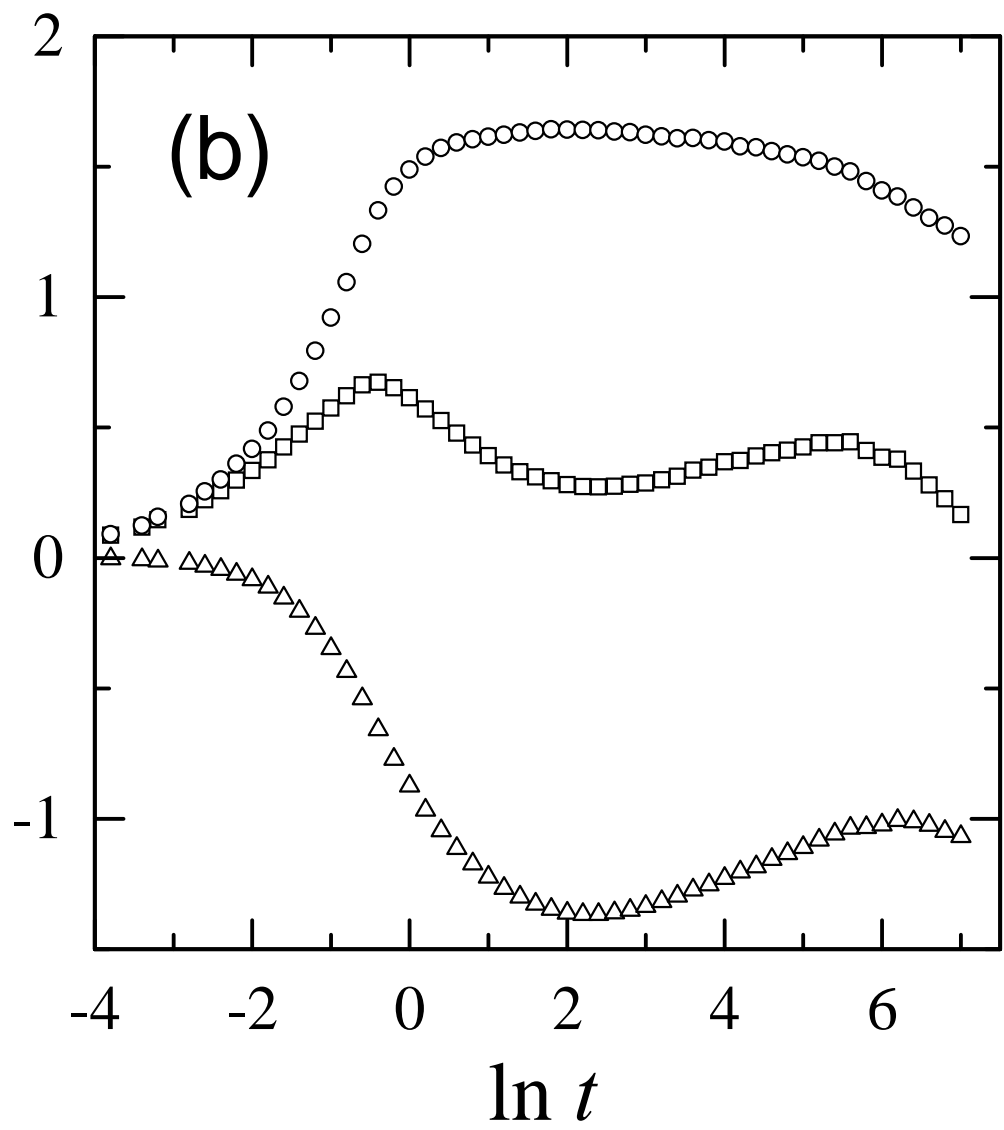
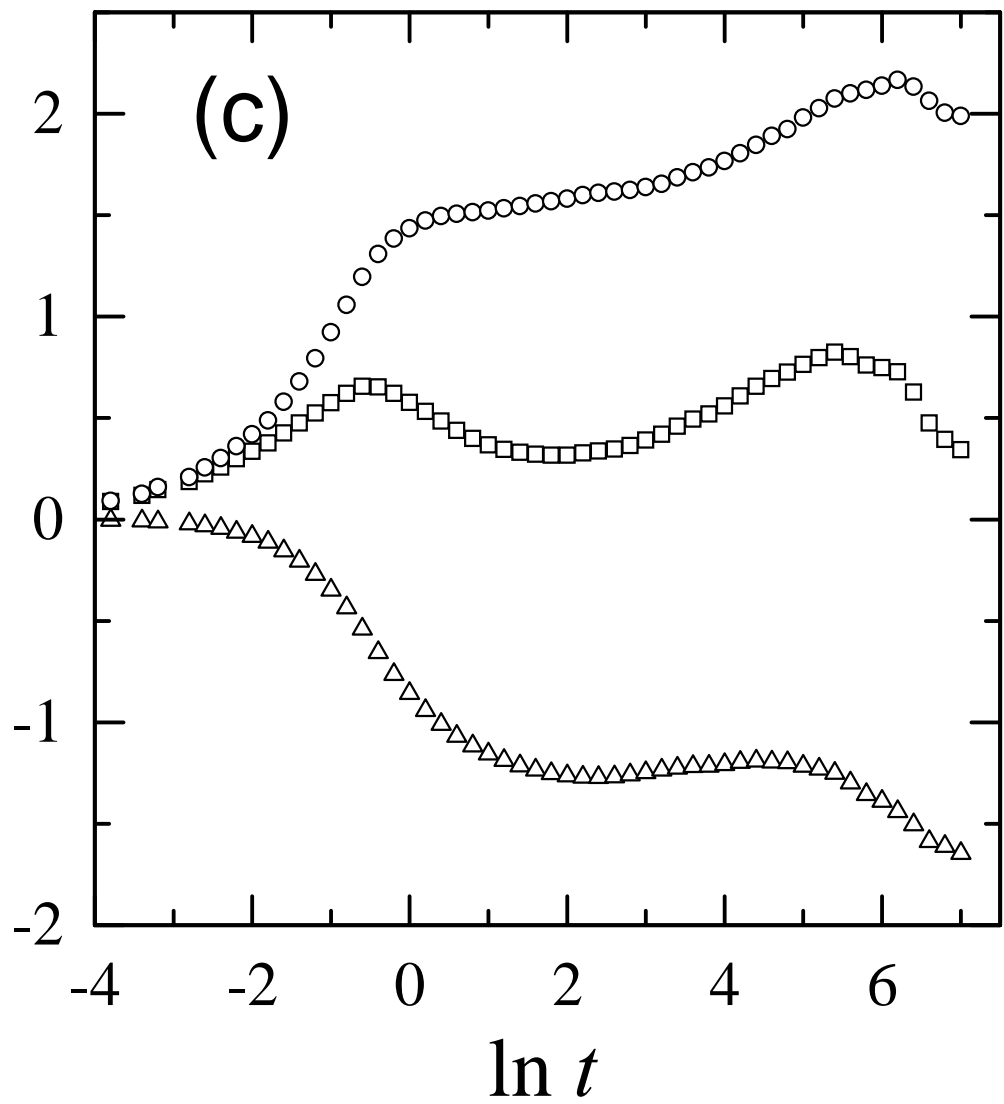
\end{document}